\documentclass[twocolumn]{aastex62}
\usepackage[utf8]{inputenc}
\usepackage[T1]{fontenc}
\usepackage{hhline}
\usepackage{makecell}

\newcommand{\sao}{\affiliation{Smithsonian Astrophysical Observatory, Cambridge, MA, USA}}
\newcommand{\harvard}{\affiliation{Department of Physics, Harvard University, Cambridge, MA, USA}}

\shorttitle{Bayesian Neural Networks: Machine Learning for Faraday Cup Calibration}
\shortauthors{Ahmed et al.}

\begin{document}

\title{Bayesian and Deterministic Neural Network approaches to Faraday Cup calibration and plasma parameter estimation}

\correspondingauthor{Lidiya Ahmed}
\email{lidiya\_ahmed@g.harvard.edu}

\author[0000-0002-2559-0831]{Lidiya Ahmed}\harvard
\author[0000-0002-7728-0085]{Michael L Stevens}\sao
\author[0000-0002-5699-090X]{Kristoff Paulson}\sao
\author[0000-0002-3520-4041]{Anthony W Case}\sao
\author[0000-0002-6145-436X]{Samuel T. Badman}\sao


\begin{abstract}
We describe a novel scheme for analyzing particle detector measurements when a well-calibrated, similarly instrumented spacecraft is present in a similar orbit. To prepare ground truth from measurements provided by a reference spacecraft, the method uses dynamic time warping (DTW)--a technique often used for pattern-matching in time series data. An artificial neural network (ANN) is created and trained to reproduce this ground truth from measurements at the target spacecraft. Unlike previous approaches, this procedure is insensitive to calibration errors in the target data stream, as the neural network may be trained from poorly calibrated particle spectra or even directly from low-level data in engineering units. We demonstrate a proof-of-concept by training an ANN to estimate solar wind proton densities, temperatures, and speeds from the DSCOVR PlasMag Faraday Cup, using the \textit{Wind} Solar Wind Experiment as a reference. We present both deterministic and Bayesian neural network approaches. Applications for Parker Solar Probe, HelioSwarm, and other missions are discussed. \\ \\
\end{abstract}



\section{Introduction}
\label{sec:intro}
Machine learning has been applied productively in different areas of solar physics research, such as sunspot classification \citep{Chola_2022}, solar flare forecasting \citep{Nishizuka_2017}, solar radio burst detection \citep{Zhang_2021}, geomagnetic storm prediction \citep{Domico_2022}, and thermal particle spectrum analysis \citep{Vech_2021}. The focus of this study lies in solar wind bulk flow analysis, using in situ measurements of velocity distribution functions (VDFs) by particle detectors such as Faraday Cups (FCs)\citep{Vech_2021}.

FCs are simple detectors that measure the flux of charged particles over finite ranges of energy-per-charge. In typical space research applications, FCs are configured to resolve the energy distributions of the most significant ion species over tens of graduations. Examples include the Solar Probe Cup (SPC) on board Parker Solar Probe \citep{Case_2020}, the Deep Space Climate Observatory (DSCOVR) PlasMag cup \citep{Loto'aniu_2022}, the \textit{Wind} Solar Wind Experiment(SWE) cup \citep{ogilvie_1995}, and many others. In this study, Solar wind proton measurements from the DSCOVR PlasMag cup and \textit{Wind} SWE cup are used. Both of these spacecraft orbit near the Lagrange 1(L1) point which is located 1.5 million kilometers from the Earth in the direction towards the Sun. The DSCOVR PlasMag cup closely resembles the \textit{Wind} SWE Cups with the distinction being that DSCOVR's circular collector plate is divided into three independent 120° sectors \citep{Loto'aniu_2022}, whereas the \textit{Wind} SWE cups each have two semi-circular collector plates. The segmentation helps to capture deflections of solar wind and understand the geometry of the plasma flow. The DSCOVR and \textit{Wind} FCs were also built in the same lab with nearly identical mechanical and electronics design (apart from the segmentation mentioned above), and they have similar energy range, measurement cadence, and operating modes. 

These instruments are calibrated in space and on the ground through careful procedures that finely characterize the sensitivity of collecting elements, the measurement circuit, and the high voltage element that provides energy discrimination. These calibrations must be monitored and updated over time with limited on-board calibration systems, through comparison with other spacecraft instruments, and with physics-based tests. 

Plasma parameters like density ($n$), thermal speed ($w$) or Temperature ($T$), and speed ($v$) are estimated by fitting the observations to a forward model of a velocity distribution function, such as a Maxwellian, using nonlinear least-squares fitting or some similar regression approach. For large data sets, nonlinear fitting can be time-consuming and computationally demanding. It may also fail to converge or improperly converge, for example when presented with unexpected distribution shapes or noisy data. This in turn can prompt ad hoc approaches and contingencies that are similarly difficult to implement with large data sets. 

This project aims to bypass the current modeling, calibration, and fitting process by implementing a robust and time-efficient approach. Artificial Neural Networks(ANNs) can help with this process by minimizing the time and effort involved in data processing.

\citet{Vech_2021} previously showed that the nonlinear regression approach can be bypassed by a well-trained ANN, increasing computational efficiency in the process. Lacking ground truth, however, the Convolutional Neural Network(CNN) in that study was trained using simulated data only for a particular range of solar wind parameters, and perfect calibration was assumed when analyzing SPC data. We take this a step further, showing that an entire data analysis pipeline segment might be similarly imitated by a well-trained ANN. We show that the raw data numbers from the instrument can be used to train an ANN without being converted to physical units. Among other advantages, this technique avoids assumptions such as perfect calibration or a particular form for the ion distribution function, making it more robust. Repeating anomalies or non-ideal instrument behaviors that are opaque to the operator might also, in principle, be learned by the ANN and mitigated. 

In this paper, we train a neural network directly with low level data, bypassing the methods employed in typical space plasma instrument pipelines. As mentioned above, \textit{Wind} SWE is used as our ground truth for training the ANN model as it provides well-calibrated and extremely well-vetted thermal solar wind ion data sets from a contemporary spacecraft in nearly the same orbit as DSCOVR. To optimize the model, we account for the time-shift difference between the two spacecraft using DTW. The DTW technique is implemented in python using the respective magnetic field measurements from both DSCOVR and \textit{Wind} as time-matching keys.  Application as a procedure for future missions such as HelioSwarm is discussed. 

\begingroup
\renewcommand{\arraystretch}{2.5}

\startlongtable
\begin{deluxetable*}{l l l}
\tablecaption{\label{Table 1} Data sources and variables for DSCOVR and \textit{Wind}. Data for each span from 2017/01/01 to 2019/06/27.}
\tablehead{\colhead{data type} & \colhead{data source} &  \colhead{variables}}
\startdata
DSCOVR PlasMag FC (level 0) & \makecell[l]{fc0 \citep{fc0}} & \makecell[l]{current spectra\\ (engineering units)} \\ 
DSCOVR PlasMag FC (science) & \makecell[l]{\detokenize{DSCOVR_H1_FC}\\ \citep{DSCOVR_H1_FC}} & \makecell[l]{\detokenize{V_GSE},\\ \detokenize{THERMAL_SPD},\\ Np} \\
\textit{Wind} SWE  & \makecell[l]{\detokenize{WI_H1_SWE}\\ \citep{wi_h1_swe}} & \makecell[l]{\detokenize{Proton_W_nonlin},\\ \detokenize{Proton_V_nonlin},\\ \detokenize{Proton_Np_nonlin}} \\
\textit{Wind} MFI  & \makecell[l]{\detokenize{WI_H0_MFI}\\ \citep{wi_h0_mfi}} & \makecell[l]{BX, BY, BZ} \\
DSCOVR PlasMag MAG & \makecell[l]{\detokenize{DSCOVR_H0_MAG} \\ \citep{dscovr_h0_mag}} & \makecell[l]{B1GSE} \\
\enddata
\end{deluxetable*}
\endgroup


\section{Data} 
\label{sec:data}
The DSCOVR science data and Solar Wind measurements from \textit{Wind} SWE used in this study were retrieved from NASA's Coordinated Data Analysis Web (NASA-CDAWeb) portal. The DSCOVR magnetic field data are from the data source "\detokenize{DSCOVR-H0-MAG}", and the variable "B1GSE" in those files contains the field vector in the Geocentric solar ecliptic (GSE) coordinate system as a function of time while the magnetic field vector components for \textit{Wind} are from "\detokenize{WI_H0_MFI}". 
The uncalibrated DSCOVR data is the faraday cup level 0 "fc0" data, which is current spectra (current as a function of instrument applied voltage) in engineering units \citep{fc0}.

The data source for \textit{Wind} SWE, is 
"\detokenize{WI_H1_SWE}" and the plasma variables in it produced by forward-modeling an anisotropic Maxwellian and fitting to the \textit{Wind} SWE Faraday Cup measurements are: proton thermal speed "\detokenize{Proton_W_nonlin}", speed "\detokenize{Proton_V_nonlin}", and density "\detokenize{Proton_Np_nonlin}". The DSCOVR uncalibrated current spectra are downloaded from the National Oceanic and Atmospheric Administration (NOAA) NEXT server. The solar wind parameters in these science data sets are derived from reduced distribution functions, which are themselves derived from calibrated current spectra acquired by the respective Faraday Cups.

A List of variables along with their data type, source, and date are shown in Table \ref{Table 1}.

\begin{figure*}[ht!]
\centering
\includegraphics[width=0.7\textwidth]{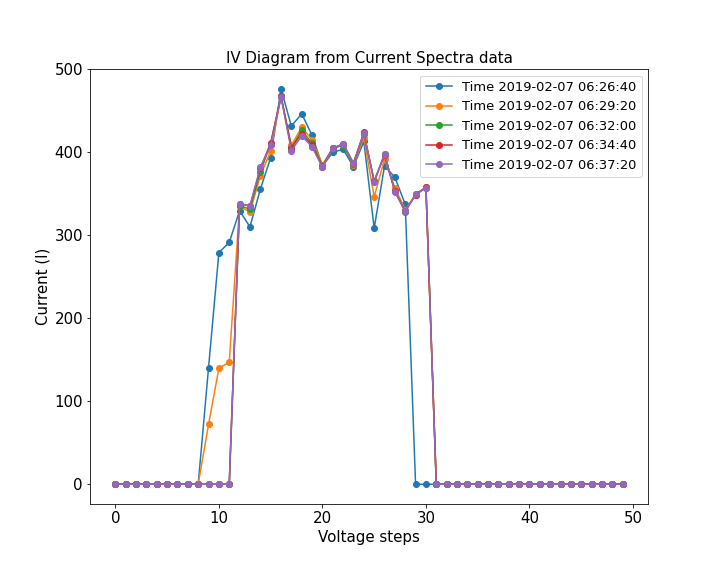}
\caption{Sample DSCOVR FC spectra data for a brief period in February 2019. Raw charge flux measurements, rescaled roughly to current, are plotted as a function of the voltage step. The ANNs are trained with current spectra as the input data, predicting solar wind parameters as an output.}
\label{Figure 1.}
\end{figure*}

The raw DSCOVR FC measurements were converted roughly to a quantity that scales with current following the procedure of \citet{dscovr_atbd} and the contributions from the instrument's three independent collector plates were summed. The DSCOVR FC acquires spectra of charged current as a function of energy by stepping through sub-ranges of a 64-element high-voltage discriminator table. For this analyses, we initially standardize the spectra by casting each as a 64-element array indexed by the voltage step, padding the unmeasured steps with zeros. We observed that most of the high- and low- extremes of the table are unused, so we truncated the array to the fifty steps that are most frequently nonzero. 

Figure \ref{Figure 1.} illustrates what the `raw' DSCOVR data look like before training the ANN model. In this figure, current as a function of voltage discriminator step is plotted. Note that, in this example, the part of the table that was visited by the experiment happens to have been concentrated in the step 8-30 range. To standardize the data set, the current is filled with zeros over the part of the voltage step table that is not visited in any given spectrum. The measured subranges suggest a background or true zero-level that is well above zero and possibly variable with respect to the voltage step, but no attempt is made to correct for these effects in the preparation. Note also that units are not present. This is because in this work \emph{the actual calibration of the DSCOVR FC is ignored}. In this data set, the abscissa is not specified, and an uncalibrated data mnemonic is used to rescale the y-value with charge flux in a roughly linear fashion. Furthermore, no background subtraction or corrections of any kind were performed. At this level of data processing, the solar wind proton peak signatures often aren't even apparent on visual inspection, which is the case in Figure \ref{Figure 1.}.

Thus prepared, the DSCOVR FC spectra and the DSCOVR magnetic field vectors were segmented and averaged down to a regular 1-minute time grid. The plasma and magnetic field measurements used in this project were already merged in the \textit{Wind} SWE H1 data set, which is provided at the SWE Faraday Cup's cadence. 

The \textit{Wind} and DSCOVR data had different sampling frequencies, and the resample function in the python library \verb+pandas+ \citep{McKinney.2010} was used for measurement frequency/cadence matching of the time-series. Both the \textit{Wind} and DSCOVR data were downsampled into a wider time-frame (160s). 
\textit{Wind} data has six columns: density, temperature, speed, bx, by, bz while DSCOVR data has 53 columns: bx, by, bz and 50 columns of current spectra. 

The DSCOVR science data are best fit parameters from the traditional method which is nonlinear fitting of velocity distribution functions. Later in this study, we will compare science data with the ANN predictions.

A period from January 1, 2017 to June 27, 2019 was chosen for this study because DSCOVR was in its prime mission during this time. Observations were relatively standard and continuous, and the \detokenize{DSCOVR_H1_FC} nonlinear fit proton parameters are available throughout as a benchmark for the machine learning method. The period provides a representative range of solar wind conditions at 1 Astronomical Unit (AU), albeit near solar minimum, over many Carrington Rotations and several orbital periods of both spacecraft. This is also a subset of the mission phase studied by \citet{Loto'aniu_2022}.

The python jupyter notebooks and data which reproduce the figures included in this paper can be found in a GitHub repository accompanying this paper. See \citep{github_repo_1}. Additionally, accompanying this paper is a reusable github repo that one can use for any two spacecraft(that are within the correlation scale), see \citep{github_repo_2}.  

\section{Methods} 
The goal of this project is to train a neural network to predict the solar wind parameters at DSCOVR from the low-level data with accuracy and precision rivaling a complete calibration and peak-fitting pipeline. The key premise is that, because the solar wind plasma is highly correlated on the spacecraft separation scales, the \textit{Wind} SWE parameters can be prepared in such a way as to provide adequate ground truth for the training process.

\begin{figure*}[ht!]
\centering
\includegraphics[width=0.7\textwidth]{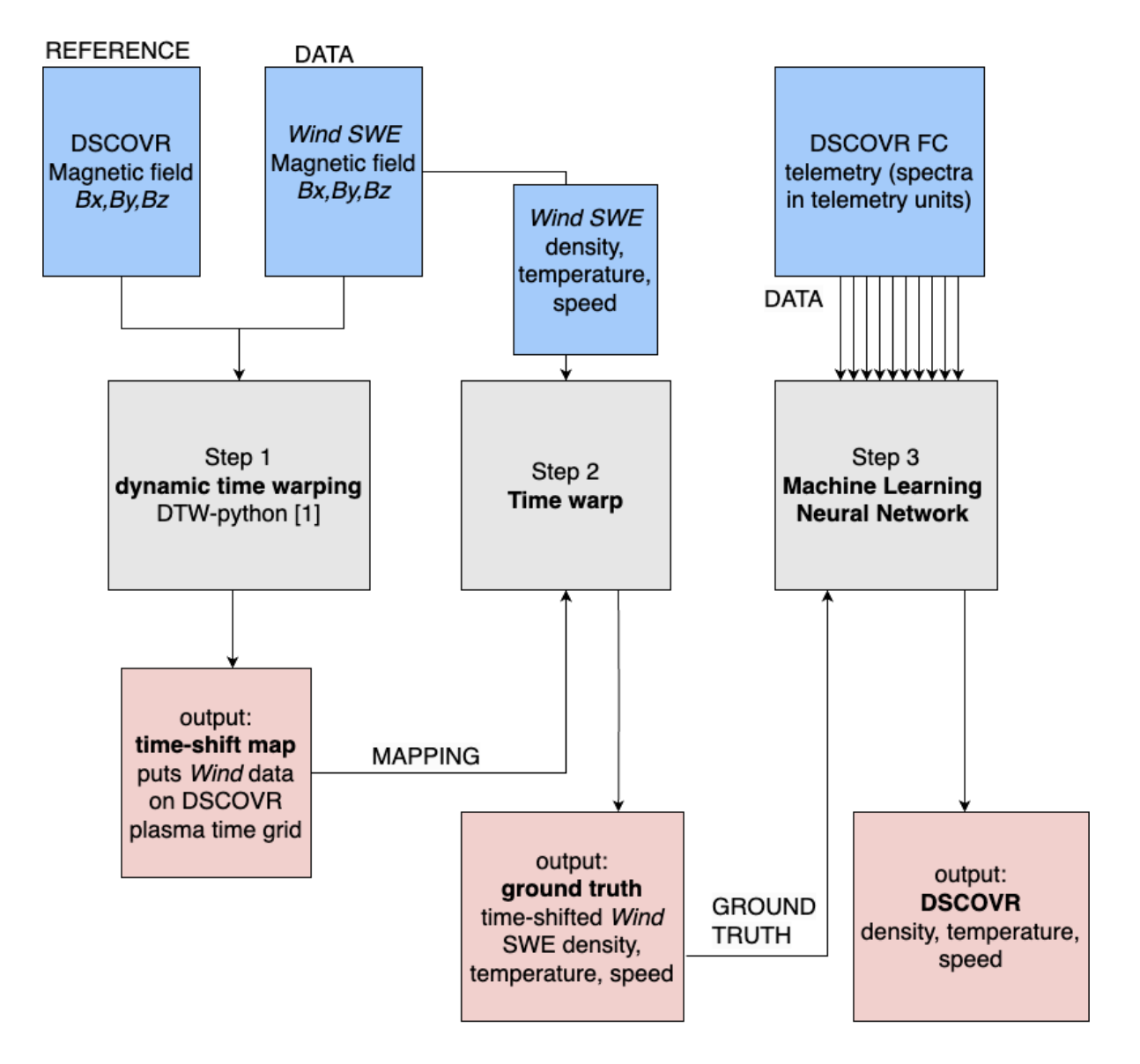}
\caption{Flow chart for the data pipeline. The well-calibrated reference (\textit{Wind}) is time-adjusted using magnetometry and used as ground truth for training the target experiment (DSCOVR FC). Note that the target experiment inputs are [re-formatted] uncalibrated telemetry.
}
\label{Figure 2.}
\end{figure*}

The methodology, as outlined in Figure \ref{Figure 2.}, comprises three major steps: (1) Deriving a DTW mapping for \textit{Wind} and DSCOVR magnetic field data, (2) Preparing ground truth by time-shifting  \textit{Wind} solar wind data using the DTW mapping from step (1), and (3) Training an ANN to reproduce the ground truth. These steps outlined in the flowchart in Figure \ref{Figure 2.} are discussed in detail below.

\cite{King_2005} analyzed solar wind observations from \textit{Wind} and the Advanced Composition Explorer (ACE), examining differences in magnetic field and plasma parameters as a function of orbital separation. The study's results show that the scale sizes $\lambda$, defined as the distance over which the autocorrelation function decreases by 10\%, are as follows: $\lambda_v \sim 1400 R_{E}$; $\lambda_{|B|} \sim 450 R_{E}$; $\lambda_{\log N} \sim \lambda_{\log T} \sim 300 R_{E}$; $\lambda_{\log Bx} \sim \lambda_{\log By} \sim 200 R_{E}$; and $\lambda_{Bz} \sim 100 R_{E}$ (also measured by \citet{Matthaeus&Dasso_2005} as $\sim 186 R_{E}$. The $x,y,z$ subscripts here denote vector components in the Geocentric Solar Ecliptic (GSE) frame.

\begin{figure*}[ht!]
\centering
\includegraphics[width=0.4\textwidth]{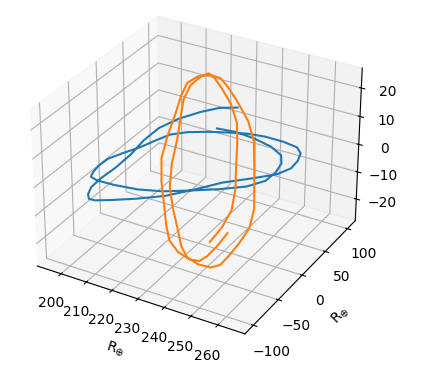}
\caption{Contemporaneous L1 orbits of \textit{Wind} (blue) and DSCOVR (orange), in GSE coordinates and in units of Earth radii ($R_{E}$).}
\label{Figure 3.}
\end{figure*}

Inspection of the \textit{Wind} and DSCOVR orbits during the DSCOVR mission shows that the spacecraft separations are less than $\sim 120 R_{E}$, as can be seen in Figure \ref{Figure 3.}. Thus, the findings of \citet{King_2005} may reasonably be expected to apply on \textit{Wind}-DSCOVR separation scales.

Furthermore, according to \cite{Loto'aniu_2022}, the correlation coefficients ($\rho$) between \textit{Wind} and DSCOVR real-time parameters were 0.94, 0.96 and 0.88 for $B_x$, $B_y$, and $B_z$ GSE, respectively. The values for the solar wind velocity were 0.96 for the GSE $v_x$ component, 0.84 for density, and 0.62 for temperature, with the expected trend of higher correlation at smaller spatial separations. For each quantity, these correlation coefficients imply that measurements from one spacecraft can predict the measurements at the other with some base degree of accuracy. Assuming perfect measurements, that accuracy would be given by $\frac{1}{2}(E/\sigma)^2 = 1-\rho$, where $E$ is the prediction error and $\sigma$ is the [natural] standard deviation of the measurable. For example, if a time series of the proton density is recorded at \textit{Wind} with a standard deviation of 5 /cc, which is typical of the solar wind at 1 AU, the correlation coefficient reported by \cite{Loto'aniu_2022} would imply that this time series also predicts the DSCOVR density to within an accuracy of about ±2.8/cc. We take these correlation coefficients and the predictive accuracy they imply as benchmarks, and we will show in the results section that higher correlations are achieved with the neural network algorithm than with the NOAA real-time pipeline.

\begin{figure*}[ht!]
\centering
\includegraphics[width=0.8\textwidth]{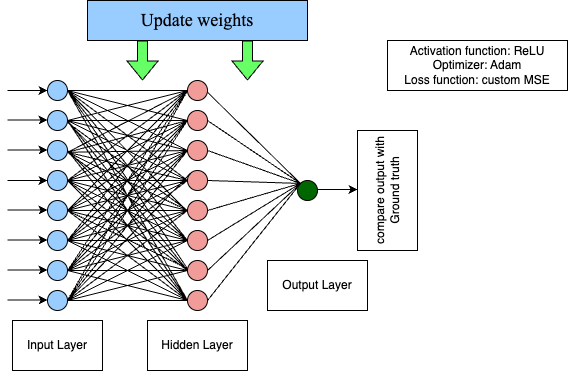}
\caption{Heuristic diagram of the Artificial Neural Networks architecture. A network of an Input Layer with input neurons in blue, Hidden layer with neurons in pink, Output Layer with an output neuron in green.}
\label{Figure 4.}
\end{figure*}

These findings show that the target measurements should be strongly correlated with candidate ground truth. We will further optimize the correlation by accounting for the variable time that it takes for plasma parcels or plasma fronts to get from one point to the other. The first two steps in the procedure are used to prepare the ground truth for the neural network by time-shifting the well-calibrated reference \textit{Wind} and putting it on the target's (DSCOVR's) time grid. The third step is to then train the neural network. The architecture of the neural network used in this study is shown in Figure \ref{Figure 4.}.

\begin{figure*}[ht!]
\centering
\includegraphics[width=0.8\textwidth]{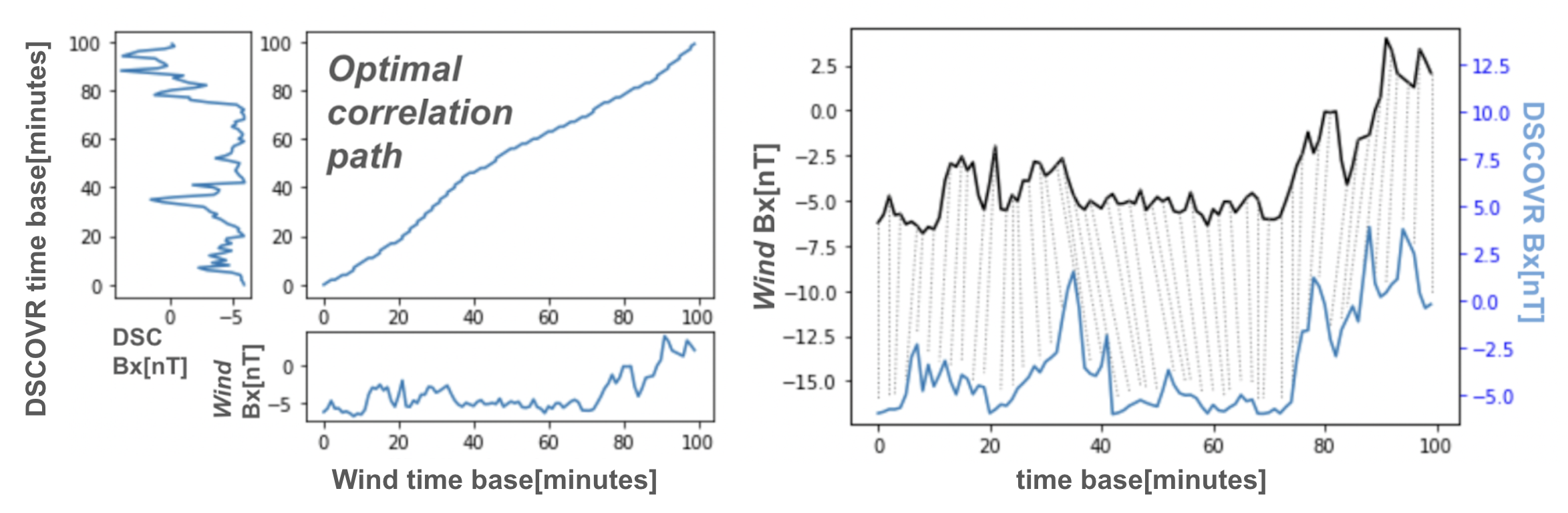}
\caption{Left illustration of DTW between a segment of magnetic field time series data from two nearby spacecraft. Right: the data sets are shown with connections indicating the derived DTW mapping. }
\label{Figure 5.}
\end{figure*}

The first step in this procedure is performed with a DTW algorithm. 
The DTW package in Python takes the time series that needs to be transformed, for which we here choose the \textit{Wind} Magnetic Field Investigation (MFI) $B_x$ component, and finds the mapping producing the best match with the reference time series, here the DSCOVR PlasMag $B_x$. This mapping includes both shifting and stretching/compression of the time series. The optimization is performed using recursion formulas that are unique to the chosen \textit{step-pattern}-- that is, the algorithm that governs and constrains the variations explored. This procedure is illustrated in Figure \ref{Figure 5.}. The set of plots in the left panel of Figure \ref{Figure 5.} shows time series segments of $B_x$ from DSCOVR along the y-axis and \textit{Wind} along the x-axis, along with the DTW mapping that optimizes the correlation between the two. The right panel shows the two [unwarped] $B_x$ time series segments on the same axes, with grey lines showing the point-by-point association between the two under the optimal mapping. The outputs of the DTW python implementation are (1) the cumulative distance metric between the warped time series and the reference and (2) either the mapping itself or the warping function. The optimal alignment was calculated using the 'dtw' function while the indexing required for the next step was created using the 'warp' function.

For this project, the DTW mappings were created using the \textit{Wind} MFI and DSCOVR PlasMag magnetic field data as time-matching keys. This is a natural choice for proving the concept because most spacecraft carrying plasma instrumentation in space are also instrumented with magnetometers and, as stated previously, the correlation in $B$ is very high for nearby pairs of spacecraft. Furthermore, the magnetic field measurements are comparatively fast, high accuracy, and high precision, to the extent that measurement uncertainties can be neglected for our purposes.

The second step in this procedure is time-shifting the \textit{Wind} SWE plasma measurements, by applying the DTW mapping from the first step, thereby creating a strong set of predictions for the plasma properties at DSCOVR. The time-warped \textit{Wind} data are then used as the ground truth for model training, and the imperfection of these predictions is accepted as the epistemic error of the training.

The final step is training an artificial neural network (ANN) to reproduce ground truth from the uncalibrated DSCOVR plasma data. In an ANN, groups of multiple neurons combine together form layers, and multiple layers are established by the sequence and interconnectedness between them. The neurons themselves are matrices of linear weights, which are adjusted through training, paired with a predetermined nonlinear activation function. In this model, there are three layers: the input layer is the interface that passes the initial data vector to the neural network for processing, the hidden layer comprises the trained neurons themselves and performs the bulk of the computations, and the output layer re-dimensionalizes to the prediction space. See Figure \ref{Figure 4.}. The input layer's dimension is 50, as each time point consists of an uncalibrated 50-element DSCOVR plasma spectrum. The output dimension is one, corresponding to the target bulk plasma parameter ($n$,$v$, or $w$).

\begin{figure*}[ht!]
\centering
\includegraphics[width=0.8\textwidth]{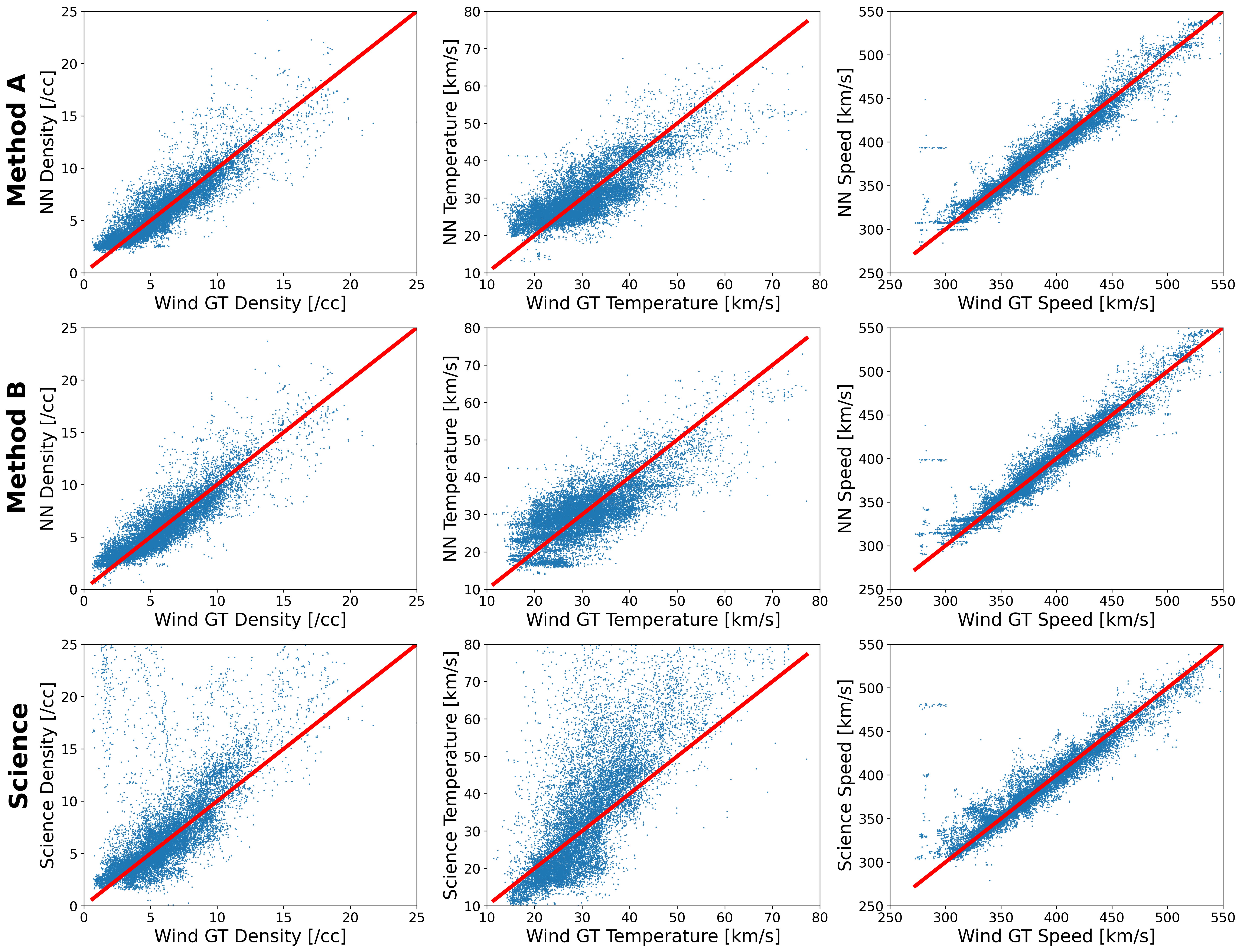}
\caption{Performance of the ANN on the April 2019 data. Top row: scatter plots of the ANN prediction for each solar wind parameter(using method A which is prepared by training on a full year of observations iteratively month by month) versus the \textit{Wind} ground truth parameter. Middle row: scatter plots of the ANN prediction for each solar wind parameter(using method B which is prepared by training initially over Jan-Feb of 2019 and then predicting solar wind parameters for the rest of the year with no further training) versus the \textit{Wind} ground truth parameter. Bottom row: scatter plots of the DSCOVR science data analysis versus the \textit{Wind} ground truth parameter. Each column represents density[/cc], temperature[km/s], speed[km/s] respectively.}
\label{Figure 6.}
\end{figure*}

ANNs are made to learn by comparing their output with the ground truth and creating a feedback loop where the weights in the network neurons are adjusted over many iterations in order to make minimize the difference. The mean squared error, which is the loss function used in this project, is the average squared difference between the output and the ground truth. Minimizing this quantity, which we refer to simply as ``loss,'' is the goal of the ANN training.

Before training the ANN, two additional steps are required: standardization to scale the data into a standard format, and splitting the data into training and testing to later test how good the model is. These are standard practices for doing machine learning regressions \citep{Fundamentals_of_ANNs_2022}. In this analysis, the data was standardized using the "StandardScaler" function in \verb+sklearn+ \citep{pedregosa2011scikit} python library. We divided the data by year into three separate campaigns, and for each we performed initial training on the months of January and February. For these months, 70\% of the data were randomly selected and used as training data, while the other 30\% were reserved for testing data. For subsequent months, training data were drawn exclusively from prior months of the same year, which will be explained below.

The initial ANN model, implemented using the \verb+keras+ \citep{chollet2015keras} and \verb+tensorflow+ \citep{tensorflow2015-whitepaper} python libraries, was created using the "Sequential" module in keras with two input layers stacked one after the other with eight neurons in each. 
The neuron activation function used in this project to introduce non-linearity is rectified linear unit (ReLU), which outputs the input for positive values and outputs zero for negative values. To improve convergence, the Adaptive Moment Estimation optimizer or 'Adam' is also used in the model training, which is a commonly used learning-rate optimizer in ANNs.

Three separate ANN models were created, one each for target parameters $n$, $v$, and $w$. Each of these was trained according to two different trial protocols, referred to simply as method A and method B.

\begin{figure*}[ht!]
\centering
\includegraphics[width=0.7\textwidth]{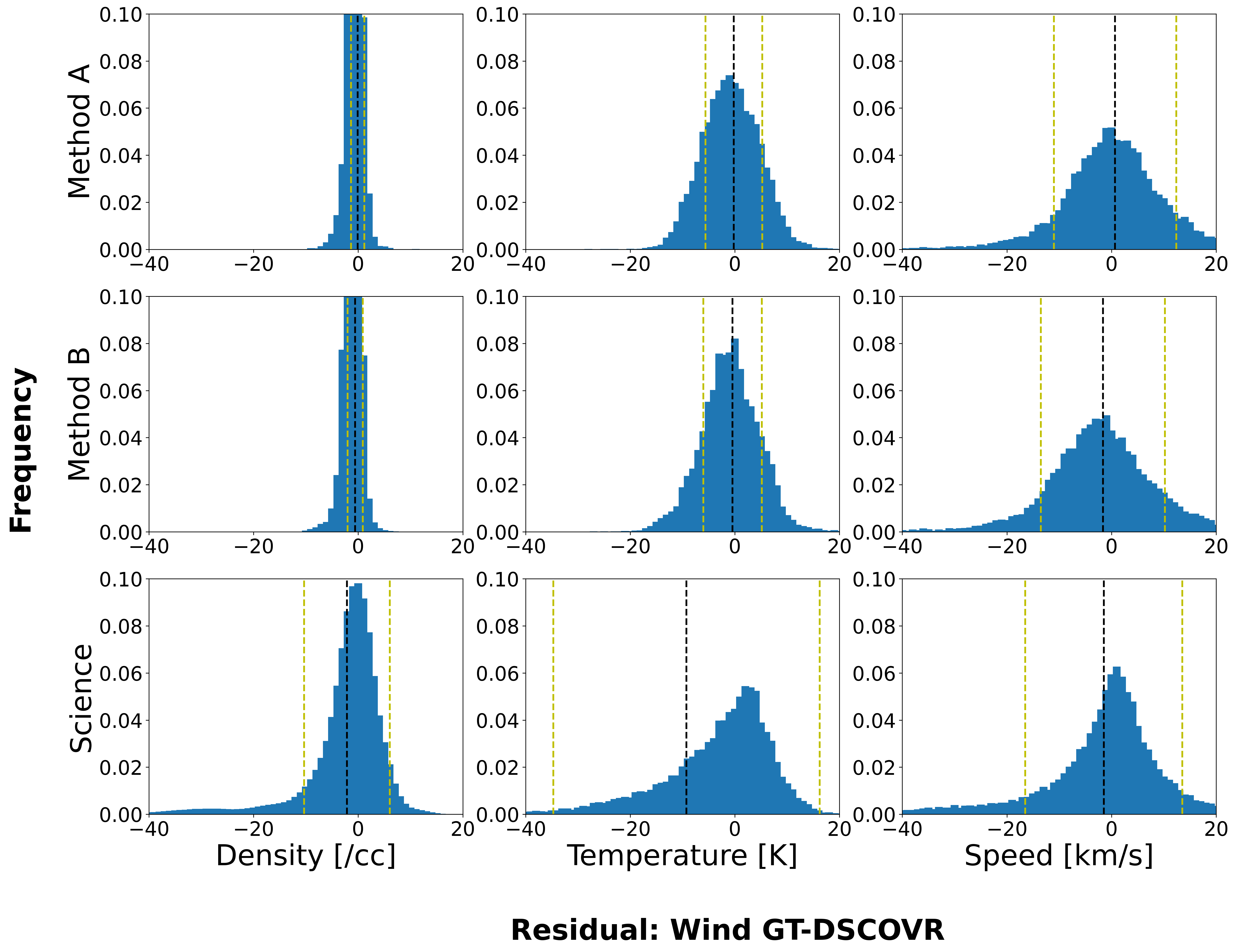}
\caption{Histograms of the residual for April 2019 where x-axis is the residuals and the y-axis is the frequency . Top row: Method A results. Middle row: Method B results. Bottom row: DSCOVR science data results.}
\label{Figure 7.}
\end{figure*}

In method A, the ANN was trained over the initial January-February period and progressively updated, month by month, to simulate a pipeline that seeks to capitalize on new ground truth as it becomes available. For example, method A predictions for the month of May are produced by training on January-February, and then performing two additional re-trainings incorporating ground truth from March and then again from April of that year.

In method B, the ANN model is prepared by training initially over January and February of a given year, and then the solar wind parameters for the rest of the year are predicted with no further training. 

The reason for trying different methods is to explore which type of training method gives the best accuracy and precision, particularly whether a fixed, stable model is effective and whether significant model skill is to be gained with updating. 

To improve the model's performance from its initial performance, updates like customizing the loss function to penalize outliers(such as negative values) were made.

In the final section, having demonstrated the effectiveness of the DTW+ANN approach, we explore the generalization to a probablistic, Bayesian Neural Network (BNN) model. In doing so, we attempt to pair plasma parameter estimation with meaningful, holistic confidence intervals and experimental uncertainties. Model confidences are validated directly against the ground truth.
This model is implemented using the 'TensorFlow Probability'\citep{(Dillon_2017)} library in \textit{keras}, while otherwise keeping the model architecture similar to the original deterministic model.

\begin{figure*}[ht!]
\centering
\includegraphics[width=0.7\textwidth]{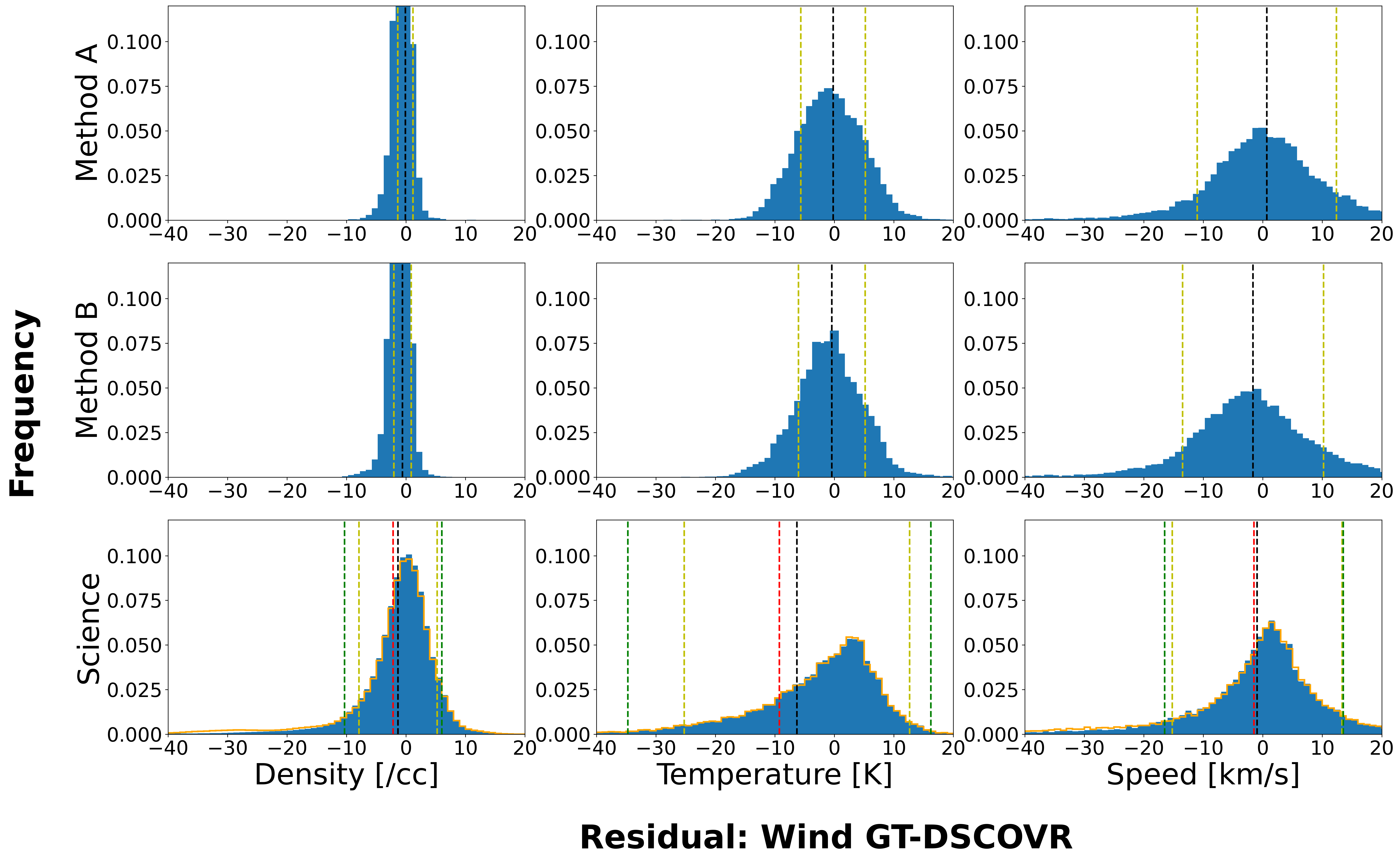}
\caption{ Histograms of the residual for April 2019 with flagged version of the science data on the right panel. Top row: Method A results. Middle row: Method B results. Bottom row: flagged DSCOVR science data results.
}
\label{Figure 8.}
\end{figure*}

\section{Results}
The results section is divided into two subsections: a case study on April 2019, and long-term analysis on all years(2017-2019) and all three parameters.  

\subsection{Case study on April 2019}
For the case study to investigate the performance of the ANN vs. traditional method, April 2019 was chosen. April provides sufficient training time before testing the predictions, and the year 2019 provides an opportunity to test the model's performance under noisy dataset conditions. 

An example of the trained ANN model performance over one selected month is presented here in depth. Each of the scatter plots in Figure \ref{Figure 6.} compares the ANN prediction and the corresponding time-warped \textit{Wind} value, denoted GT for Ground Truth, for a considered solar wind parameter and method for the month of April 2019. The top row in Figure \ref{Figure 6.} shows the predictions over the course of the month employing Method A: trained in this case initially on Jan-Feb 2019 data and then re-trained to include March 2019. The middle row shows similar plots for a model trained using method B: only one training on Jan-Feb 2019 data. As a benchmark, the last row compares the publicly available DSCOVR science data products to the same shifted \textit{Wind} data.

The ANN scatter plots are consistent upon inspection with a linear relationship between the ANN predictions and ground truth across the full domain, with a degree of scatter that will be considered in depth shortly. The science data products show comparable scatter, providing an initial indication that the DTW+ANN technique may be capable of precision comparable to experimental uncertainties. Groups of outliers are apparent in the bottom row where the DSCOVR science data set significantly overshoots the shifted \textit{Wind} data for some measurements of density and temperature, particularly when the shifted \textit{Wind} measurement is low. These outliers must indicate either that DSCOVR has observed a set of dense and hot features not observed at \textit{Wind} or that the DSCOVR science data analysis has produced poor fits to the low level data. These overshoots are absent in the ANN predictions, which could be consistent with the former hypothesis. If the latter hypothesis is true, however, it would imply that the ANN is more skilled than the science algorithm under certain conditions. Inspection of the \detokenize{DSCOVR_H1_FC} quality flag variable supports the latter hypothesis: approximately 8\% of the science data for this period have been marked as having low or unknown fit quality, and the flagged measurements present disproportionately with high density and temperature.

Of the three physical parameters, speed is estimated with the highest precision by all methods, as can be seen in the scatter plots in Figure \ref{Figure 6.}. This is expected; given that the input is a low level representation of the ion distribution function and that the speed corresponds directly to the solar wind peak location in energy, it is the most straightforward parameter to accurately and precisely measure. Density and temperature involve the respective integrated amplitudes and widths of the peaks, which are more sensitive to noise and non-Maxwellian forms of the distribution. As \citet{Loto'aniu_2022} and \citet{ogilvie_1995} both note, the speed measurement is significantly more accurate than either density or temperature for these instruments. 

Temperature is the least precise ANN prediction. One distinction between the DSCOVR and \textit{Wind} Faraday Cups is that the latter capitalizes on the rotation of the \textit{Wind} spacecraft in order to measure two anisotropic proton temperature components, whereas DSCOVR does not rotate. Based on the empirical record of temperature anisotropies at 1 AU \citep{Wilson2018}, estimates based on Faraday Cup measurements from a non-spinning spacecraft are limited to $\sim$10\% precision in thermal speed due to anisotropy effects. Plenty of scatter would therefore be expected here for even a perfect model.

Residual histograms for this period are shown in Figure \ref{Figure 7.}, illustrating the relative accuracy and precision, with respect to the ground truth, for these three estimation methods. In this case, the ANN models produce symmetrical, normal distributions in the residual parameter. The distributions are centered very nearly at zero, corresponding to high overall accuracy, as compared to the science data residuals. 
The average accuracy (peak center) values of the residual histograms for April 2019 as shown in Figure \ref{Figure 7.} were, method A: 0.90/cc for density, 4.09 km/s for temperature, and 7.92 km/s for speed, method B: 0.96/cc for density, 4.08 km/s for temperature, and 8.10 km/s for speed, and science data: 3.07/cc for density, 16.72 km/s for temperature, and 8.82 km/s for speed. 
They are also significantly narrower in this example, indicating that the ANNs also achieved higher precision. The average precision (peak width) values were, method A: 1.29 /cc for density, 5.28 km/s for temperature, and 11.57 km/s for speed, method B: 1.37 /cc for density, 5.25 km/s for temperature, and 11.98 km/s for speed, and science data: 7.31/cc for density, 23.00 km/s for temperature, and 15.01 km/s for speed. 
We also note that the histograms for the science data residuals are skewed and asymmetric, having long tails that indicate a tendency to overshoot the ground truth.

\begin{figure*}[ht!]
\centering
\includegraphics[width=0.7\textwidth]{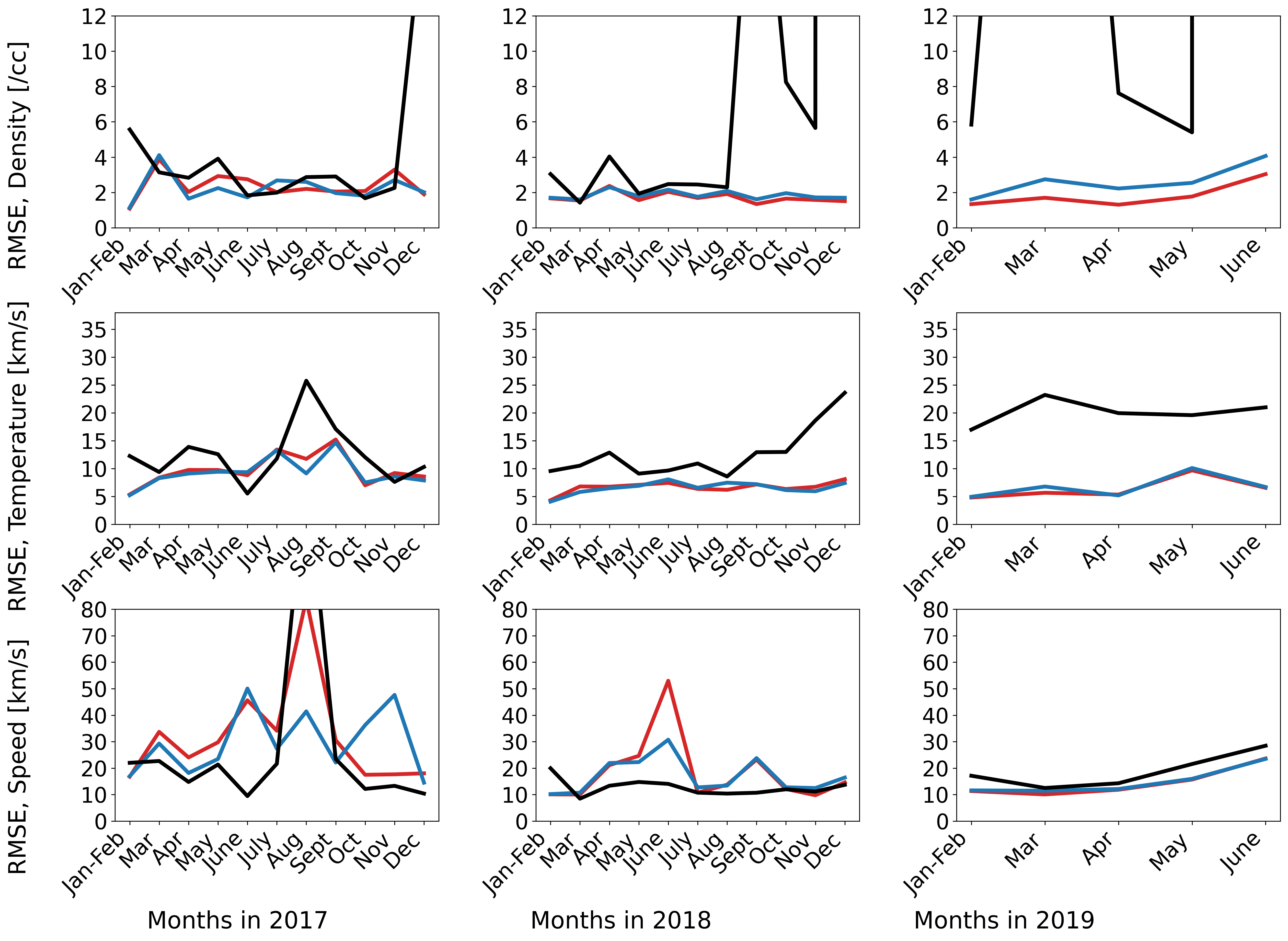}
\caption{The month-by-month precision of two neural network models that have been trained to predict time-shifted \textit{Wind} SWE solar wind measurements from DSCOVR block spectra. Model A (red) is iterative training, Model B (blue) is only initial training(Jan-Feb), and the precision of the DSCOVR science data analysis (black) is also shown. After having trained the ANN on measurements from Jan-Dec 2017, it is further trained with with data from a subsequent period of time: 2018 and 2019. 
}
\label{Figure 9.}
\end{figure*}

It is interesting to ask whether the ANN rivals the more traditional data pipeline even when flagged, low-confidence fits are removed from the DSCOVR science data. To address this, we repeat the comparison, but redact measurements that were categorized in the \detokenize{DSCOVR_H1_FC} metadata as ``low confidence on the peak,'' ``suspiciously long disagreement with \textit{Wind},'' or ``noisy data or spikes.'' As can be seen in Figure \ref{Figure 8.}, this filter improves the agreement between the DSCOVR and \textit{Wind} science data, shifting the centers of the residual histograms to nearly zero, reducing the peak widths, and somewhat reducing the overshoot tail. The residual histograms are still skewed, however, and significantly broader than the ANN versions. For this month of observations, both ANN methods appear to deliver comparatively higher accuracy, precision, and completeness while avoiding outliers.

\subsection{Isolating sources of error}

What remains to be determined is whether the ANN has missed true differences between the solar wind conditions at \textit{Wind} and DSCOVR. It is important to consider the possibility that, for this interval, the ground truth preparation and the DSCOVR measurements \textit{should not} agree due to some set of real phenomena local to one spacecraft. As previous studies have shown, the solar wind is highly correlated on these scales, so such phenomena should be rare. To support the assertion, we inspect the two magnetic field time series for significant structural differences. After investigating the magnetic field data for both spacecraft, it was found that the outliers in the DSCOVR data were spikes that did not also appear in the magnetic field data, suggesting that they are errors rather than real phenomena. 

We also investigated whether the existing quality flag in the DSCOVR science data may be insufficient to remove all unphysical fits to the low level data. From visual inspection of the time series, it is evident that the DSCOVR science data contain significantly more spikes than the corresponding \textit{Wind} plasma data and there is no natural reason for this to be so.

To better compare with the DSCOVR science data under conditions where that data pipeline has been effective, we applied a [generic] spike filter algorithm to the DSCOVR science data and restricted the comparison to exclude suspicious fits. The effect was significant. For example, when these outliers were removed from the DSCOVR science data set, the density uncertainty estimate over 2017 was adjusted to 4.65\%, down from 9.02\%. The revised figure is comparable to the ANN error estimate of 4.79\%.

When averaged by year across 2017 and 2018 after spike filtering, uncertainty estimates for the DSCOVR science density dropped to 5.96\%, temperature to 11.48\%, and speed 4.52\% overall, which are respectively 1.52\% higher, 2.71\% higher, and 1.39\% lower than the corresponding ANN uncertainties.

Therefore, after applying the spike filter, the DSCOVR science data pipeline would seem to match the performance of the ANN to within $\sim$3\%. This supports the hypothesis  that the difference is indeed in data processing rather than real physical differences. 

We interpret the spikes as originating from bad peak fits converging upon artifacts in the low level DSCOVR data. This is consistent with known issues in the dsocvr science pipeline, which lead us to expect an abundance of artifacts in the low-level data from which the DSCOVR science data are derived \citep{Loto'aniu_2022} \citep{dscovr_atbd}.This is also evident from the fact that the DSCOVR science data quality flags are set much more frequently than the \textit{Wind} SWE \citep{DSCOVR_flag} \citep{Wind_flag}.

It is also informative to ask what fraction of the error is attributable to the ANN fitting method by itself. Therefore, we applied the ANN to \textit{Wind} current spectra and trained it following the same protocol as the DSCOVR ANN, obviously minus the DTW ground truth preparation step. Because the instrument operations are not exactly the same (they use different voltage tables, \textit{Wind} spins and DSCOVR does not, DSCOVR samples a bit faster), this was not a strictly "apples-to-apples" comparison. Nonetheless, the sunward-most set of data frames from one of the \textit{Wind} SWE cups during full-range scans, which are acquired every 15 minutes, provide a very similar raw data spectrum to the DSCOVR analogue.

This training provides a believable proxy for the baseline fitting error from the ANN.  The results are included in Table \ref{tab:Table 3}. When comparing the \textit{Wind} ANN to DSCOVR ANN, we find that: 

\begin{enumerate}
\item The ``baseline'' fitting error for density was 4.81\%. This is actually slightly higher than the uncertainty estimate from the DSCOVR ANN, but by a difference of 0.38\%. We assert that this difference is not significant, and most likely reflects the aforementioned differences in the data preparation. 
\item The ``baseline'' fitting error for temperature was 6.24\%, as compared to the DSCOVR ANN estimated uncertainty of 8.78\% (post outlier removal).
\item The ``baseline'' fitting error for speed was 3.62\%, as compared to the DSCOVR ANN error of 5.91\%. 
\end{enumerate}

We learn from this that the spike removal exercise resolves the major part of the discrepancy between this new method and the DSCOVR science data pipeline, and that the remaining discrepancy is comparable to or slightly larger than the uncertainty inherent to fitting these data with this model. The third and final contribution to the discrepancy, which is physical differences in the plasma that are not well-described by DTW, appears to be smaller than either of these. This is as expected, given the strong correlation between the magnetic field measurements at the two spacecraft.

\subsection{Long-term study}
After initially training the ANN on measurements from Jan-Dec 2017, it is further trained with data from a subsequent period of time: 2018 and 2019. Results for all years(2017-2019) and all three parameters are shown in this subsection.

In addition to the scatter and residual plots, the ANNs also showed smaller Root Mean Squared Error (\% RMSE) values than the traditional method, whose variation over time is shown in Figure \ref{Figure 9.}. In this figure, the RMSE values of method A, method B, as well as science data are shown in red, blue, and black respectively. 
The top row in this Figure \ref{Figure 9.} shows density plots, while the middle row shows temperature plots, and the bottom row shows speed plots for all training years. The left column is for 2017, while the middle column is for 2018, and the right column is for 2019. It can be concluded from the results that the ANN predictions in both methods have lower RMSE values than DSCOVR science data except for 2017 and 2018 speed data where the ANN is slightly outperformed by the traditional method, while still producing very comparable results. It can also be observed in general that the iterative training method (method A) does not seem to improve on the initial training (method B) as expected. There might be several reasons that can explain this, such as catastrophic forgetting by the ANN--forgetting previous training information when learning new information.

\startlongtable
\begin{deluxetable*}{l l l l l l l }
\tablecaption{ \label{tab:Table 2} Correlation coefficient and measurement uncertainty values for ANN vs. traditional method.}
\tablehead{\colhead{Parameter} & \colhead{Year} &  \colhead{DSCOVR science {$\rho$}} & \colhead{DSCOVR ANN {$\rho$}}& \colhead{DSCOVR science E\%} & \colhead{DTW+ANN E\%} & \colhead{No DTW ANN E\%} }
\startdata
Speed & 2017 & 0.94 & 0.93 & 5.98 & 7.00 & 7.02 \\
Speed & 2018 & 0.98 & 0.97 & 3.37 & 4.82 & 5.31 \\
Density & 2017 & 0.80 & 0.81 & 9.02 & 4.79 & 5.23 \\
Density & 2018 & 0.69 & 0.87 & 14.23 & 4.09&  4.53 \\
Temperature & 2017 & 0.80 & 0.84 & 12.33 & 9.46& 9.73 \\
Temperature & 2018 & 0.77 & 0.90 & 13.10 & 8.09& 8.66 \\
\enddata
\end{deluxetable*}

To explore the limits of the model, correlation coefficient({$\rho$}) values between the ground truth(\textit{Wind}) and DSCOVR were calculated using the numpy 'corrcoef' function, which computes the Pearson correlation coefficient between two arrays. The proportional percentage RMSE values were also calculated using $\%RMSE = RMSE / $range$(Wind) * 100$ where RMSE values were calculated as the square root of the values returned by the $'mean\_squared\_error'$ function in sklearn library. The measured values in Table \ref{tab:Table 2} imply that the ANN produces better {$\rho$} values than the traditional method and that \textit{Wind} can predict DSCOVR parameters to better than ±10\% which is the target range for space weather forecast/applications.

We also investigated how much DTW improves the ANN performance relative to an approach without DTW. For a perfect model, we would expect the \%RMSE values for the ANN without DTW  to be closer to \%RMSE for science data than the RMSE values for ANN+DTW models. But what we observed here as can be seen in Table \ref{tab:Table 2} is that the ANN+no DTW measurements are closer to ANN+DTW measurements than traditional/science measurements. Which implies that DTW is indeed making improvements, but not as significant as expected. The improvement values, averaged by year were 3.90\% for 2017, 8.50\% for 2018, and 34.90\% for 2019. When averaged by parameter, the percentage values were: 22.09\% for Density, 8.53\% for Temperature and 16.65\% for Speed. The DTW's highest improvements were observed in noisy data, i.e 2019 data. 

Overall, these results underscore the ANN's capability to predict solar wind parameters with high precision.

\startlongtable
\begin{deluxetable*}{l l l l l l }
\tablecaption{ \label{tab:Table 3} Measurement uncertainty values for DSCOVR ANN, \textit{Wind} ANN vs. spike-filtered traditional method.}
\tablehead{\colhead{Parameter} & \colhead{Year} &  \colhead{DSCOVR science E\%} & \colhead{Filtered DSCOVR science E\%} & \colhead{DTW+ANN E\%}  & \colhead{\textit{Wind} ANN E\%} }
\startdata
Speed & 2017  & 5.98 & 5.66 & 7.00  & 3.69\\
Speed & 2018  & 3.37 & 3.37 & 4.82  & 3.54\\
Density & 2017 & 9.02 & 4.65 & 4.79 &  5.17\\
Density & 2018  & 14.23 & 7.28 & 4.09&   4.46\\
Temperature & 2017  & 12.33 & 10.36 & 9.46 & 6.15\\
Temperature & 2018 & 13.10 & 12.61 & 8.09&   6.32\\
\enddata
\end{deluxetable*}

\begin{figure*}[ht!]
\centering
\includegraphics[width=0.7\textwidth]{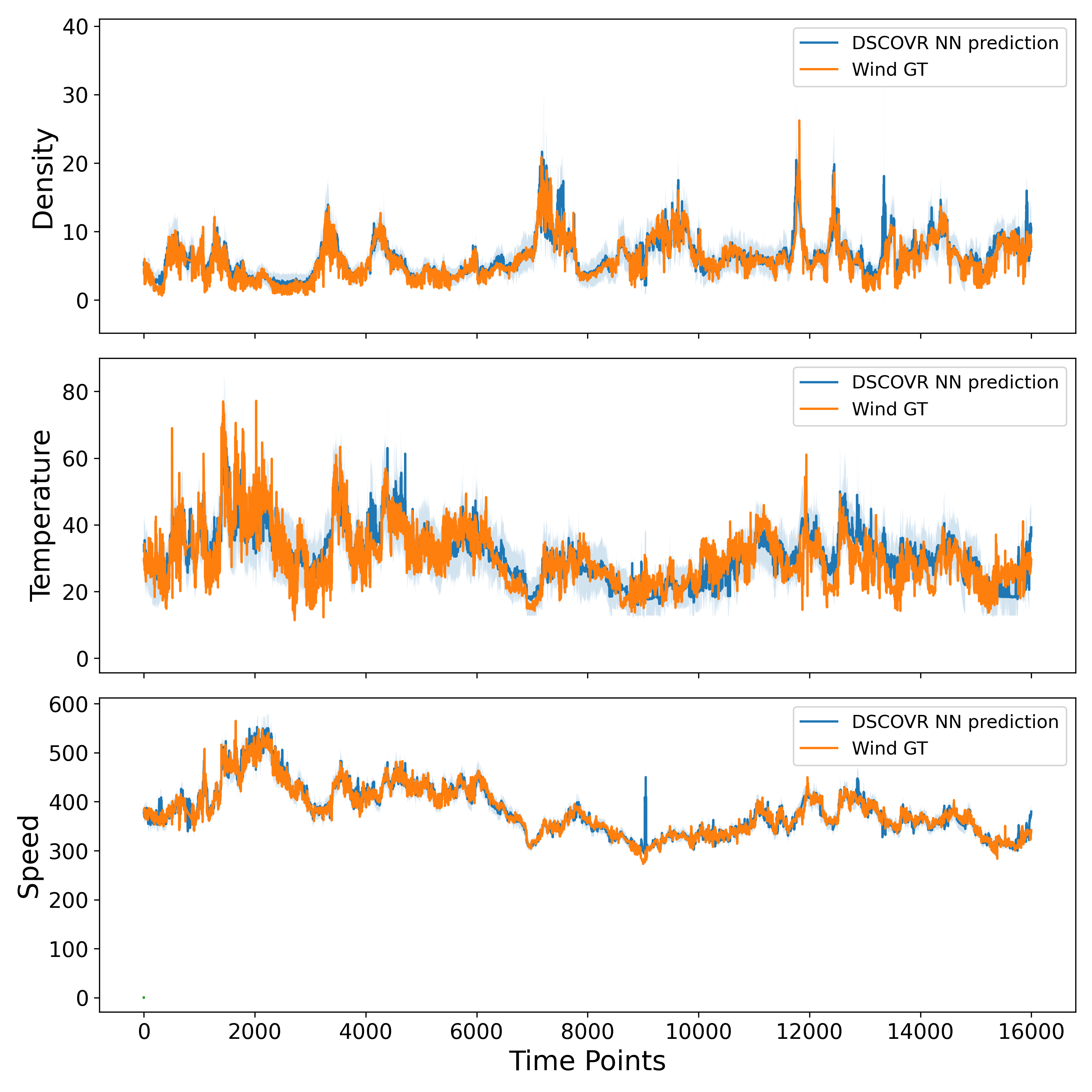}
\caption{ 95\% Confidence Interval plots of the ANN on April 2019 data. The panels are density, temperature, speed as a function of time point in minutes, from top to bottom. The confidence interval(CI) is shown in grey, ANN prediction in blue, \textit{Wind} ground truth in orange. 
}
\label{Figure 10.}
\end{figure*}

The last step in this project was generalizing the ANN to a Bayesian Neural Network(BNN) model using the 'TensorFlow Probability' library in keras. BNNs produce a range of possible outcomes with associated uncertainties instead of single point estimates, which is the case for standard deterministic models. This final step is very important in measuring uncertainties and assigning levels of confidence to predictions. 

\begin{figure*}[ht!]
\centering
\includegraphics[width=0.7\textwidth]{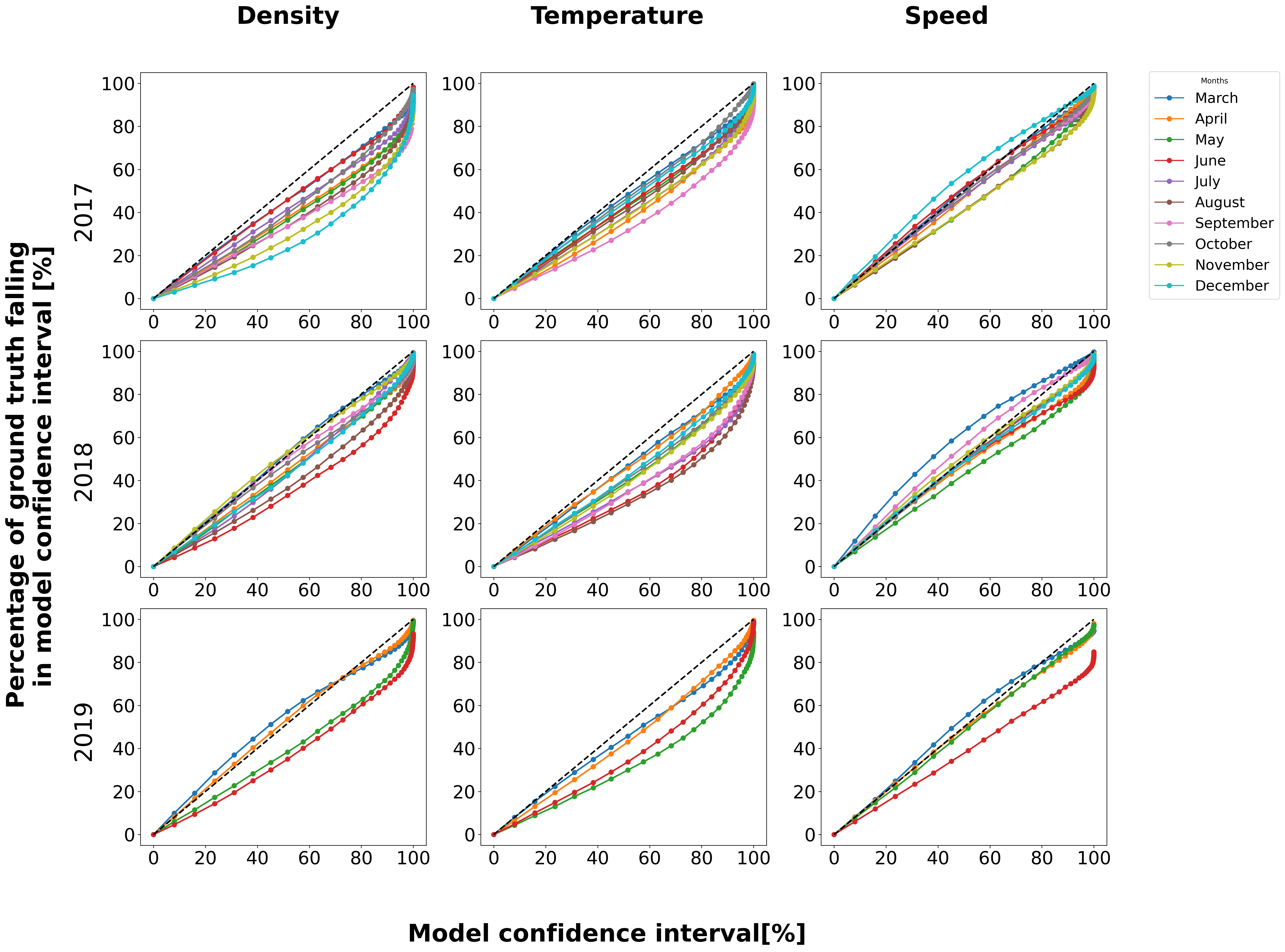}
\caption{ Calibration curve: The percentage of ground truth falling in model's confidence interval vs the model's confidence interval. A linear increase suggests the model is well-calibrated and its confidence intervals are accurate. All months are plotted in different colors, for years 2017-2019 and for each parameter. }
\label{Figure 11.}
\end{figure*}

The BNN learns weight distributions instead of specific weights as in the case of deterministic neural networks (DNN). Therefore, we define prior and posterior distributions of the weights according to Baye's theorem, where the goal in the training process is to learn the parameters of these weight distributions such as mean and variance. For the BNN  model in this project, a normal distribution with mean=0 and stddev=1 was used as a prior: an initial belief about the weights before seeing any data, while the posterior was defined as a normal distribution with learnable/variable mean and stddevs where the goal is to optimize them.

The BNN's overall architecture was similar to the DNN except the BNN being modeled as distributions. Further investigation of different ML techniques revealed the modifications that optimized the model best, such as using softplus--a smoother version of ReLU activation function, customizing the loss function to penalize the model for underestimation of peaks(a persistent problem), and making the priors trainable instead of fixed.


A sample 95\% confidence interval(CI) for the BNN model is shown below in Figure \ref{Figure 10.} where the CI is shown in grey. The narrow CI range shows that the model is very confident about its predictions and that its predictions are in good agreement with the ground truth. But in order to validate the CI plots, the percentage of ground truth falling in model's confidence interval was plotted against the model's confidence interval for all years and all parameters, as shown in Figure \ref{Figure 11.}. The overall linear increase observed in this figure suggests the model is well-calibrated and its CIs are accurate. 

The slight shift from the identity line in these plots indicates model over- or under-confidence. Overconfidence, which is more typical here, means that a smaller fraction of ground truth values fall in the model's confidence interval than expected. This implies that the model's CI intervals are too narrow and/or that relatively extreme outcomes are underrepresented in the model. Overconfidence is most pronounced in the temperature estimate, which also generally carries the highest absolute uncertainty. In the case of density and speed, the model confidence is more accurately estimated, but there are still deviations from perfect CI calibration, usually tending towards overconfidence. 

The overconfidence effect could be because the BNN model weights are normally distributed, while natural differences in the data set more likely follow subtler statistical distributions with power-law tails. Overall, the confidence intervals are verified to within about 10\%, and they are clearly a useful estimate of the true measurement uncertainty. For comparison, the conventional 1-sigma error bar corresponds to the 68th percentile confidence interval of the model. This validation exercise suggests that the BNN gets that error bar right to within about 25-50\%-- one might expect this model as trained to overconfidently quote a 10 km/s error on some particular speed estimate, for example, when the true measurement uncertainty is closer to 15 km/s.

To implement this as a data processing pipeline, simply re-normalizing according to the calibration curve (such as rescaling the confidence intervals) might be a good first step. But, since this is systematic, a more robust approach as an extension of this project would be to explore re-scalings or different model weight distribution approaches to fine tune. Updating the priors to distributions that capture real-world variability (with heavy tails), for example, using re-calibrating techniques, could improve the BNNs uncertainty estimation.

The model is overconfident in certain regions and underconfident in others. Our hypothesis about the overconfidence is that it is likely due to the assumption that the errors are normally distributed when, in reality, they may have a more heavy-tailed distribution. Underconfidence could also occur, for example, in regions where the model has not been very well trained on, leading it to be less confident and overestimate its uncertainty. The systematic overconfidence in temperature predictions could be due to higher variability in temperature data 
which the assumed priors in the BNN did not capture. It could also be the poorer correlation between nearby spacecraft for temperature reported by \citep{King_2005}. 

The main point here is that the BNN’s assumption of Gaussian uncertainty might not fully capture the actual more heavy-tailed data distribution. This suggests that a BNN model that accounts for this behavior better might improve the estimation of uncertainty.

Lastly, the 1-sigma error bars between the latest BNN models and the DNNs were compared as can be seen in Table \ref{tab:Table 4}. We can see from these values that the \%RMSE values between the BNN and DNN models are similar, matching performances with the added advantage of having uncertainties included in the BNN models' predictions.

\startlongtable
\begin{deluxetable*}{l l l l l l }
\tablecaption{ \label{tab:Table 4} percentage RMSE for DNN vs BNN.}
\tablehead{\colhead{Parameter} & \colhead{Year} &  \colhead{Method B DTW+DNN E\%} & \colhead{Method A DTW+DNN E\%}& \colhead{E\% BNN} & \colhead{E\% science} }
\startdata
Speed & 2017 & 6.44 & 7.0 & 6.33 & 5.98  \\
Speed & 2018 & 4.53 & 4.82 & 5.02 & 3.37  \\
Speed & 2019 & 5.13 & 4.99 & 5.23 & 6.45\\
Density & 2017 & 4.75 & 4.79 & 5.47 & 9.02  \\
Density & 2018 & 4.58 & 4.09 & 4.41 & 14.23  \\
Density & 2019 & 7.43 & 5.33 & 6.65 & 17.73 \\
Temperature & 2017 & 9.15 & 9.46 & 9.60 & 15.25 \\
Temperature & 2018 & 7.96 & 8.09 & 9.26 & 13.10 \\
Temperature & 2019 & 9.12 & 8.67 & 9.40 & 27.21 \\
\enddata
\end{deluxetable*}

\section{Discussions and Conclusions}
This study has been intended as a demonstration for an end-to-end calibration and analysis system that can easily integrate new information from a well-calibrated reference, i.e predict solar wind moments from DSCOVR PlasMag Faraday Cup current spectra measurements, using \textit{Wind} SWE measurements as a reference. In solar wind research, accurate measurements of these solar wind properties is important leading to improved models of solar wind and better prediction of space weather events. 
Our computationally inexpensive implementation uses two input layers stacked one after the other (with eight neurons in each). Three separate ANN models were then trained to predict measurements for solar wind temperature, density and speed separately. Figure \ref{Figure 9.} shows a comparison of the RMSE values of the ANN models trained with data from 2017, 2018, and 2019 in two different training methods as well as the traditional method. The plot shows that the ANN is predicting the solar wind parameters with lower RMSE values than the traditional method. The scatter plots and residual plots in Figures \ref{Figure 6.}, \ref{Figure 7.}, and \ref{Figure 8.} also show that the ANN model has a better performance than the traditional method. Additionally, further investigation of DTW's contribution to the improvement of the ANN models show that DTW does indeed make improvements, even though not as significant as expected. Finally, the BNN models match the performance of the initial DNN models with an added benefit of modeling the uncertainties in the predictions.  

In the future, we will expand to a larger volume of training data conditions, benchmark performance against conventional inter-spacecraft calibrations (such as OMNIweb \citep{Papitashvili_King_2020, King_2005}) etc. The procedure could be applied to conjunctions at scales where the correlation between nearby spacecraft observations remains strong, such as Parker Solar Probe commissioning near Earth \citep{Fox_et_al_2016}, or to the calibration of new L1 observatories like Interstellar Mapping and Acceleration Probe (IMAP) \citep{IMAP}. 

This method could be even more instrumental for calibration and data analysis in tight-formation multi-spacecraft missions like HelioSwarm \citep{helioswarm}, a NASA-funded Medium-Class Explorers (MIDEX) currently in development. HelioSwarm is a constellation of 9 spacecraft, with 8 nodes and 1 hub, which is designed to study turbulence in the solar wind. The fundamental measurements of that mission will be the [typically very small] temporal and spatial variations in plasma parameters on small scales upstream of Earth. Mission success will depend critically on having detailed, uniform calibrations across the entire swarm. With a conventional approach, instrumenters on that mission would have to perform a calibration process that involves fine-tuning a predetermined calibration table or model on orbit for the each detector in order to harmonize between the different spacecraft observations. They would also have to perform the usual forward-model fits to the plasma spectra in order to derive plasma parameters and estimate errors. The protocol demonstrated in this paper, however, could simply train the response of each node to the ground truth established at the hub over a sufficient period.

Significantly, in a multi-spacecraft mission scenario all scientific data would also serve as calibration data, facilitating ongoing improvement and adaptation as the mission encounters a diverse range of conditions. This adaptability would allow the swarm calibration to learn and respond to new conditions promptly as they present at the hub, ultimately enhancing the mission data quality.

In conclusion, we have demonstrated a proof of concept for a plasma instrument calibration and analysis system that predicts solar wind properties directly from raw current spectra, without requiring expert knowledge of instrumentation or physical meaning of the raw data. Using this method, we achieved absolute accuracies on the density, temperature, and speed of the plasma that are superior to the traditional pipeline and analysis. We achieved measurement precisions comparable to or better than the conventional data analysis from the DSCOVR Plasmag Faraday Cup experiment, and also comparable to the verifiable limit from multi-spacecraft intercalibration. We also find evidence that this method is more robust against anomalous conditions and/or fitting errors that occur in the traditional data pipeline, such that this method may result in broader derived data coverage in some cases. Finally, we have shown that holistic and verifiable measurement uncertainties are derived with this approach. 

The approach demonstrated here can be readily be applied alongside traditional calibration and data analysis to improve outcomes for science and space weather missions with plasma diagnostic components. It will be particularly powerful for constellation mission scenarios, such as HelioSwarm, or for new missions that commission or orbit in the vicinity of a well-calibrated predecessor, such as L1 space weather monitors.

\section{Acknowledgement} 
This work was supported by NASA DSCOVR grant 80NSSC20K1845 and NASA \textit{Wind} data analysis grant 80NSSC20K0647. The author also acknowledges valuable feedback from Prof. Douglas Finkbeiner and Prof. Carlos Arguelles, and the support of the Harvard physics department. 
Results from the data analysis presented in this paper as well as a github repository with the python jupyter notebooks used for data analysis are directly available from the authors. See \citep{github_repo_1} \citep{github_repo_2}.
DSCOVR and \textit{Wind} data have been accessed through the NASA SPDF Coordinated Data Analysis Web(CDAWeb) and the NOAA next server.

\bibliography{main}

\end{document}